\documentclass[conference]{IEEEtran}
\IEEEoverridecommandlockouts
\usepackage{cite}
\usepackage{amsmath,amssymb,amsfonts}
\usepackage{bm}
\usepackage{algorithmic}
\usepackage{graphicx}
\usepackage{textcomp}
\usepackage{xcolor}
\usepackage{mathtools}
\usepackage{nicematrix}
\usepackage{multirow}

\usepackage[colorlinks=true, allcolors=blue, pdfborderstyle={/S/U/W 1}]{hyperref}

\newcommand{\blockcomment}[1]{}
\newcommand{\blc}[1]{{\color{black}#1}}

\newcommand{\infoclear}[1]{{\color{black}#1}}
\newcommand{\acc}[1]{{\color{black}#1}}

\def\BibTeX{{\rm B\kern-.05em{\sc i\kern-.025em b}\kern-.08em
    T\kern-.1667em\lower.7ex\hbox{E}\kern-.125emX}}
\begin{document}

\title{Evaluating the resolution of AI-based accelerated MR reconstruction using a deep learning-based model observer
\blockcomment{*\\
{\footnotesize \textsuperscript{*}Note: Sub-titles are not captured in Xplore and
should not be used}
\thanks{Identify applicable funding agency here. If none, delete this.}
}}

\author{
\IEEEauthorblockN{Zitong Yu\textsuperscript{1}\thanks{\textsuperscript{1}\text{Now at Mayo Clinic (e-mail: \href{mailto:yu.zitong@mayo.edu}{yu.zitong@mayo.edu}).}}}
\IEEEauthorblockA{\textit{DIDSR/OSEL/CDRH} \\
\textit{US Food and Drug Administration}\\
Silver Spring, USA \\
\href{mailto:Zitong.Yu@fda.hhs.gov}{zitong.yu@fda.hhs.gov}}
\and
\IEEEauthorblockN{Rongping Zeng}
\IEEEauthorblockA{\textit{DIDSR/OSEL/CDRH} \\
\textit{US Food and Drug Administration}\\
Silver Spring, USA \\
\href{mailto:rongping.zeng@fda.hhs.gov}{rongping.zeng@fda.hhs.gov}}
\and
\IEEEauthorblockN{Frank Samuelson}
\IEEEauthorblockA{\textit{DIDSR/OSEL/CDRH} \\
\textit{US Food and Drug Administration}\\
Silver Spring, USA \\
\href{mailto:frank.samuelson@fda.hhs.gov}{frank.samuelson@fda.hhs.gov}}
\and
\IEEEauthorblockN{\hspace{7.5cm}Prabhat Kc\textsuperscript{2}\thanks{\textsuperscript{2}\text{Corresponding author.}}}
\IEEEauthorblockA{\hspace{7.5cm}\textit{DIDSR/OSEL/CDRH} \\
\textit{\hspace{7.5cm}US Food and Drug Administration}\\
\hspace{7.5cm}Silver Spring, USA \\
\href{mailto:prabhat.kc@fda.hhs.gov}{\hspace{7.5cm}prabhat.kc@fda.hhs.gov}}
\and
\blockcomment{
\IEEEauthorblockN{Nirmal Soni}
\IEEEauthorblockA{\textit{DIDSR/OSEL/CDRH} \\
\textit{US Food and Drug Administration}\\
Silver Spring, USA \\
\href{mailto:more@fda.hhs.gov}{Nirmal.Soni@fda.hhs.gov}}
\and
\IEEEauthorblockN{And more}
\IEEEauthorblockA{\textit{DIDSR/OSEL/CDRH} \\
\textit{US Food and Drug Administration}\\
Silver Spring, USA \\
\href{mailto:more@fda.hhs.gov}{more@fda.hhs.gov}}
}}

\maketitle
\begin{abstract}
\acc{Deep Learning-based Model Observers (DLMOs) were developed to evaluate a multi-coil sensitivity encoding parallel MRI at different acceleration factors on the Rayleigh discrimination task as a surrogate measure of resolution. Gaussian-convolved singlet and doublet signals with varying intensities and lengths were inserted into the white matter of synthetic brain images. K-space data were generated using a simulated MRI at acceleration factors of one ($1\times$, fully sampled), $4.9\times$, and $16.4\times$, and reconstructed using a conventional root-sum-of-squares (rSOS) method and an AI-based U-Net method. DLMOs were first trained on fully sampled images and then fine-tuned for each acceleration factor using transfer learning. With a human-label alignment training strategy, the DLMOs achieved discrimination performance similar to that of trained human observers. Resolution was assessed using the area under the receiver operating characteristic curve (AUC), while PSNR and SSIM provided complementary task-agnostic comparisons. Although the U-Net method yielded significantly higher PSNR and SSIM than rSOS across different acceleration factors (p$<$0.05), task-based evaluation using the proposed DLMO showed inferior performance relative to fully sampled reconstruction. U-Net ($4.9\times$) exhibited modest gains over rSOS ($4.9\times$) for short signals (4–5 mm), but its AUC decreased by approximately 25\% and 5\% for 4 mm and 5 mm signals, respectively, compared with rSOS ($1\times$). Similar declines were observed for U-Net ($16.4\times$). These results demonstrate that AI-based accelerated MR reconstruction may improve visual appearance, \infoclear{but may not preserving task performance}. The proposed \infoclear{DLMO approach} may be employed to characterize the discriminative efficacy of AI-based undersampled MRI reconstruction.}
\end{abstract}


\section{Introduction}
Accelerated magnetic resonance (MR) imaging has shown great promise for reducing acquisition time and improving patient comfort. Techniques such as parallel imaging and compressed sensing exploit undersampling in k-space to shorten scan duration while attempting to maintain image quality \cite{deshmane2012parallel,hamilton2017recent}. More recently, artificial intelligence (AI)-based reconstruction methods have emerged, offering visually appealing results, even at high acceleration factors \cite{knoll2020deep}. Despite these advances, smoother reconstructions by those AI-based reconstruction algorithms may come at the expense of diagnostic information \cite{mcdonnell2003data}, particularly the fine details \infoclear{typically needed} for making clinical accurate decisions. The efficacy of those AI-based accelerated MR reconstructions cannot be purely assessed from their visual appearance.

Metrics like peak signal-to-noise ratio (PSNR) and structural similarity index (SSIM) have long been used to compare MR images reconstructed by AI-based methods with their reference standards \cite{sriram2020grappanet,aggarwal2018modl}. Although convenient and easy to compute, it is well-known that such metrics are task-agnostic and do not necessarily correlate with performance on clinically relevant tasks \cite{yu2020ai,li2021assessing, yu2023need, kc2024assessing}. In clinical practice, medical images are acquired for specific clinical tasks, such as detecting lesions or defects, quantifying biomarkers, or discriminating fine structures. Observer studies provides a methodological foundation for such task-based evaluations by quantifying performance on clinically relevant tasks. Human observer studies remain the primary practice for such evaluations, but they pose several challenges. Recruiting trained readers is costly, and large sample sizes are \infoclear{employed} to ensure statistical significance in the observed results. Besides, ground truth of clinically relevant findings (e.g., lesions) is also \infoclear{necessary} to judge the performance of observers on the task, further complicating study design. These \infoclear{constraints} make them difficult to scale, particularly when multiple MR reconstruction methods or acceleration factors are to be compared. Variability in human performance adds another layer of complexity. Readers with different levels of expertise may produce markedly different outcomes, and even the same observer may \infoclear{not} provide consistent assessments of identical images due to intra-reader variability. These considerations have led to both the use of model observers (MO) and the use of simulated images. 

Classical MOs, including Hotelling observer and channelized Hotelling observer (CHO), have shown great promise for detection tasks in medical imaging \cite{myers1987addition,wollenweber1999comparison,sankaran2002optimum,yu2024ctless}. However, they often rely on hand-crafted features and may \infoclear{not} generalize across heterogeneous imaging conditions. Recently, deep learning (DL) has introduced a new opportunity for building more flexible and adaptive MOs \cite{zhou2019approximating, li2024application, gong2019deep, fan2020deep}. Zhou et al. developed a DL-based approach to approximate ideal observer (IO) test statistics on a signal-known-exactly/background-known-exactly detection task with a simulated parallel-hole collimator system \cite{zhou2019approximating}. Li et al. applied a convolutional neural network (CNN)–based MO to estimate data-space IO performance for binary signal detection in a stylized X-ray breast CT system \cite{li2024application}. Gong et al. reported strong correlations between a DL-based MO (DLMO) and human observers in a lung nodule detection task \cite{gong2019deep}, while Fan et al. showed that a DLMO agreed more closely with human observers than either the linear non-prewhitening matched filter or the CHO \cite{fan2020deep}. By learning task-relevant features directly from data, DLMOs \infoclear{are able to} capture task-specific information from images acquired under different conditions while retaining correspondence with human observer performance.

In this study, we developed a DLMO \infoclear{approach} to evaluate AI-based reconstruction algorithms for a multi-coil sensitivity encoding (\textbf{SENSE}) parallel magnetic resonance imaging (MRI) system across different acceleration factors. We calibrated the performance of our DLMOs with human observer performance using a novel human-label alignment training strategy, and validated the similarity between DLMOs and human observers in two-alternative forced choice (\textbf{2AFC})\cite{barrett2013foundations} studies. The performance of reconstruction algorithms was evaluated on the Rayleigh discrimination task as a surrogate measure of resolution. In this task, reconstruction algorithms are typically evaluated by observers’ performance on identifying whether there is a pair of dot signals or a line signal presented in the image. However, such stylized structures are rarely encountered in human anatomy, limiting the direct applicability of patient-based studies. Although this could be potentially addressed by conducting physical phantom studies, these studies have limitations in comprehensively representing the real patient population. Further, clinical trials would \infoclear{involve} imaging the same patient multiple times across different acceleration factors. This process would be expensive, time-intensive and have multiple logical challenges. In this context, the in-silico imaging\footnote{Badano in \cite{badano2021silico} defines in-silico imaging as the computer simulation of an entire imaging system including source, object, detection, and image interpretation components used for research, development, optimization, technology assessment, and regulatory evaluation of new technology to complement bench testing .} trial paradigm provides an alternative mechanism by evaluating imaging technology rigorously and objectively in simulated clinical scenarios that address these challenges by providing the ability to simulate patient-population variability and imaging-system physics \cite{abadi2020virtual,jha2021objective,badano2018evaluation}. Moreover, \infoclear{in-silico imaging allows} the same virtual patient \infoclear{to} be scanned multiple times across multiple acceleration factors, overcoming the practical challenges in clinical trials. Motivated by these advantages, we designed an in-silico imaging trial to evaluate the performance of a conventional root-sum-of-squares (rSOS) reconstruction approach and an AI-based U-Net reconstruction approach using our \infoclear{DLMO}. We also evaluated reconstruction algorithms using fidelity-based figures of merit (FoMs), including PSNR and SSIM. Through this study, we demonstrate the applicability and feasibility of the proposed \infoclear{DLMO} for identifying reductions in diagnostic performance due to MR acceleration.

\section{Materials and Methods}

\subsection{Problem Formulation}
Consider a multi-coil SENSE parallel MRI system acquiring the spatial distribution of MR signal within a human body, denoted by a vector $f(\bm{\mathfrak{r}})$, where $\bm{\mathfrak{r}} \in \mathbb{R}^3$ denotes the 3-dimensional coordinates, the MRI system yields k-space measurement $\bm{g}_i \in \mathbb{C}^{M\times 1}$ for $i^{th}$ coil, where $M$ is the number of elements in the acquired k-space data. Denote the sampling mask of k-space by $\bm{\Phi} \in \mathbb{C}^{M\times 1}$, the Fourier transform by $\mathcal{F}$, the coil sensitivity map by $\bm{S}_i$ and complex-valued Gaussian noise by $\bm{n}_i \in \mathbb{C}^{M\times 1}$ for $i^{th}$ coil, the imaging acquisition process in the MRI system can be expressed by 
\begin{equation}
\label{eq:img_acq_singlet_coil}
    \bm{g}_i = \bm{\Phi}\mathcal{F}\bm{S}_i f(\bm{\mathfrak{r}}) + \bm{n}_i.
\end{equation}
Eq.~\eqref{eq:img_acq_singlet_coil} can be written more concisely as
\begin{equation}
\label{eq:img_acq}
    \bm{g} = \mathcal{H}f(\bm{\mathfrak{r}}) + \bm{n},
\end{equation}
where $\bm{g} = \{\bm{g}_1, \bm{g}_2, ... ,\bm{g}_{N_c}\}$ is the set of acquired k-space data across coils, $\bm{n} = \{\bm{n}_1, \bm{n}_2, ... ,\bm{n}_{N_c}\}$ is the set of noise associated with k-space data acquisition across coils. $N_c$ denotes the number of receiver coils, and $\mathcal{H}$ denotes a continuous-to-discrete imaging operator that maps $\mathbb{L}_2(\mathbb{R}^3) \rightarrow \mathbb{C}^{M \times N_c}$. The goal in multi-coil SENSE parallel MRI is to reconstruct $f(\bm{\mathfrak{r}})$ given the undersampled k-space measurement $\bm{g}$. Denoting the reconstructed image by an N-dimensional vector $\bm{\hat{f}}$, and the reconstruction operator by $\mathcal{R}$, we have
\begin{equation}
    \bm{\hat{f}} = \mathcal{R}\bm{g}
\end{equation}
In this study, we designed a Rayleigh discrimination task to evaluate the resolution of reconstructed images \cite{sanchez2014task,hanson1991rayleigh}. To conduct evaluations on such a task, we consider an observer to distinguish whether a singlet (a line) signal or a doublet (a pair of neighboring points) signal presented in the reconstructed image. The singlet hypothesis and the doublet hypothesis are denoted by $H_{s}$ and $H_{d}$, respectively, and can be described as:
\begin{equation}
\begin{split}
    H_{s}: \bm{\hat{f}} & = \mathcal{R} \left[\mathcal{H}\left(\bm{f_b}+\bm{f_{s_1}}\right)+\bm{n}\right], \\
    H_{d}: \bm{\hat{f}} & = \mathcal{R} \left[\mathcal{H}\left(\bm{f_b}+\bm{f_{s_2}}\right)+\bm{n}\right], \\
\end{split}
\end{equation}
where $\bm{f_b}$, $\bm{f_{s_1}}$, and $\bm{f_{s_2}}$ represent the background, singlet signal, and doublet signal functions, respectively.

To perform a Rayleigh discrimination task, an observer computes a test statistic $t(\bm{\hat{f}})$ that maps the reconstructed image $\bm{\hat{f}}$ to a real-valued scalar variable, which is then compared to a predetermined threshold $\tau$ to classify $\bm{\hat{f}}$ as satisfying $H_s$ or $H_d$.

\subsection{The Proposed Deep Learning-based Model Observer (DLMO)}
\begin{figure}[h]
\centering
\includegraphics[width=\linewidth]{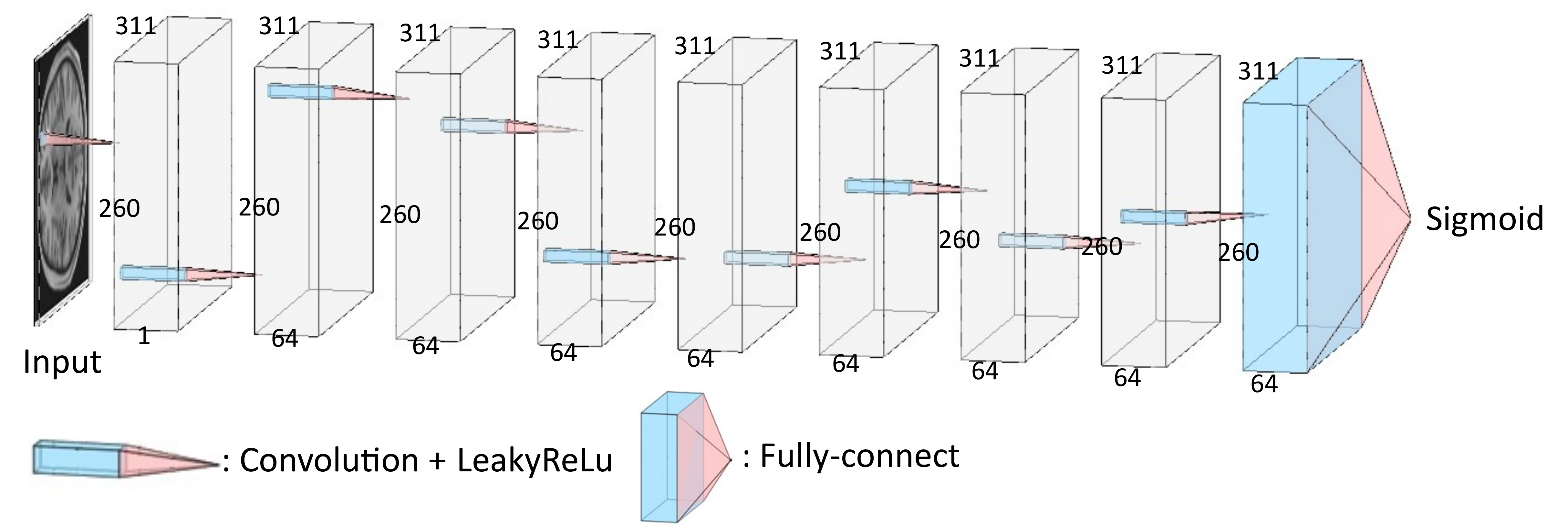}
\caption{The structure of the proposed DLMO.}
\label{fig:DLMO_strct}
\vspace{-1.0em}
\end{figure}

The architecture of the proposed DLMO comprises eight convolutional layers followed by a fully connected layer. The network input was a reconstructed image $\bm{\hat{f}}$ with dimensions $260 \times 311$, and the output was the estimated probability of satisfying hypothesis $H_d$. Each convolutional layer employed $7 \times 7$ kernels with a stride of one. The first seven convolutional layers contained 64 filters, whereas the final convolutional layer included a single filter. A leaky rectified linear unit (LeakyReLU) activation function was applied after each convolutional operation. A fully connected layer using a sigmoid activation function produced the final probability estimate. The architecture of the proposed DLMO is shown in Fig.~\ref{fig:DLMO_strct}. We consider the \infoclear{scalar} value prior to the sigmoid activation function as the test statistic $t$ yielded by the DLMO and \infoclear{use} it to classify $\bm{\hat{f}}$ as satisfying $H_s$ or $H_d$.

\subsection{In-silico Experiment Setup} 
\begin{figure}[h]
\centering
\includegraphics[width=\linewidth]{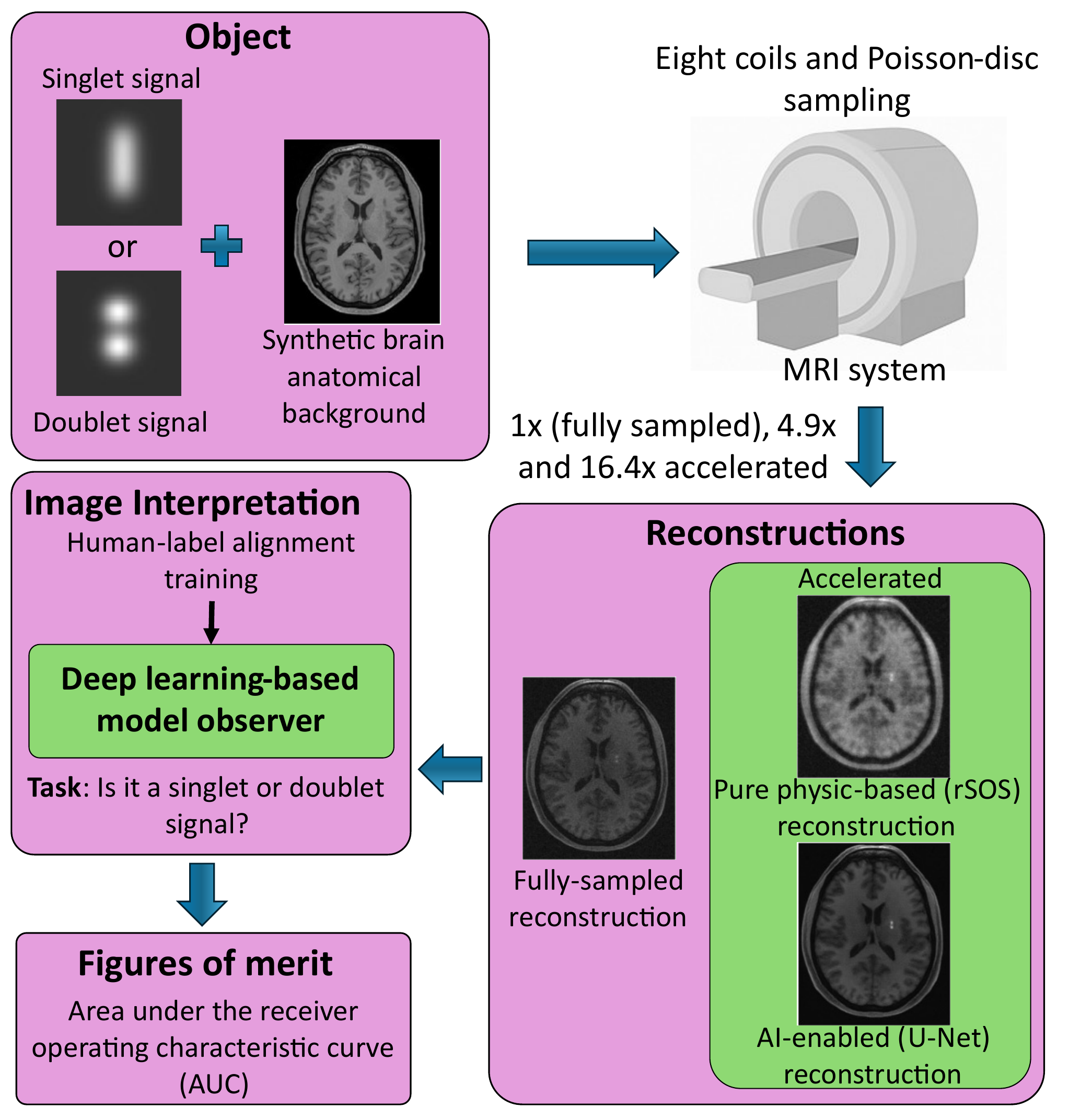}
\caption{The schematic of the evaluation workflow.}
\label{fig:framework}
\vspace{-1.0em}
\end{figure}

We designed an in-silico imaging trial to evaluate the resolution of accelerated MR reconstruction methods using the proposed DLMO. Brain MR scans were generated using a simulated multi-coil SENSE parallel MRI system at three acceleration factors and reconstructed with two approaches, including the conventional physics-based rSOS method and an AI-based U-Net. Resolution of the reconstructed images was assessed with the proposed DLMO on the Rayleigh discrimination task, considering a range of signal intensities and lengths. Additional evaluations included fidelity-based FoMs and qualitative visual inspection. A schematic of the \infoclear{DLMO evaluation} is shown in Fig.~\ref{fig:framework}. Detailed descriptions of the evaluation studies are provided in the following subsections.

\subsubsection{Synthetic Brain Data Generation}\label{sec:data}
A stochastic object model (SOM) was employed to generate the to-be-imaged objects, using a deep learning-based generative model trained by Li et al. in \cite{li2025estimating}. Specifically, synthetic two-dimensional axial brain MR slices were generated using a denoising diffusion probabilistic model (DDPM).
 The DDPM was trained on axial brain MR image slices from the Human Connectome Project (HCP) dataset \cite{van2013wu,HCP_dataset}. Only slices containing the cerebrospinal fluid region were selected to train the DDPM model to reduce variability arising from slice location. The generated images had pixel size of 1 mm and their intensities were normalized to the range from 0 to 1. Image statistics, including image mean, image standard deviation,  gray and white matter ratios, estimated from the DDPM-generated images were validated to be similar to those from the HCP real brain MR images. These comparative results are provided in the supplementary material.

Because stylized structures of singlet and doublet signals are rarely encountered in human brain, we inserted synthetic signals with clinically relevant parameters. The doublet signal was a pair of neighboring points and the singlet signal was a line, both of which were convolved by a Gaussian kernel with zero mean and a standard deviation of 1.75 mm. We modeled signal intensities (maximum pixel value ranging from 0.2 to 1.9) and signal lengths (distance between the centers of neighboring points or length of the line ranging from 4 mm to 14 mm), similar to the range of non-specific white matter lesions observed in the fastMRI+ dataset \cite{zhao2022fastmri+}. The doublet and singlet signals were inserted into the white matter regions of the simulated brain images, with randomly selected locations. Each image sample contained only one signal type (i.e., either singlet or doublet).

We generated in total \acc{$433,380$ brain images using the DDPM brain model. Of those, $80,000$ brain images were used to train the AI-based U-Net reconstruction approach, $336,000$ brain images were used to train the DLMO network, $16,000$ were used as the test dataset, and the remaining $1,380$} brain images were used in the validation of DLMO with human observers.

\subsubsection{Simulated MR Acquisition System}
K-space data from synthetic objects (discussed previously in \ref{sec:data}) were acquired using a simulated 8-coiled SENSE parallel MRI system. The Berkeley Advanced Reconstruction Toolbox (\textbf{BART})\cite{tamir2016generalized} was used to perform the simulated MR acquisition.\acc{To model undersampled acquisitions, BART's Poisson-disc sampling was used with a $32 \times 32$ auto-calibration region at the center of k-space (i.e., the central $32 \times 32$  k-space region was fully sampled). The undersampling acceleration factors in the two phase-encoding directions $(k_y, k_z)$ were set to $(2, 2)$ and $(4, 4)$, resulting in effective acceleration factors of approximately $4.9\times$ ($20.6$\% of the original k-space data retained) and $16.4\times$ ($6.1$\% of the original k-space data retained), respectively. The $4.9\times$ and $16.4\times$ acceleration factors used in this study represent typical and aggressive undersampling levels, respectively, consistent with those reported in the recent literature for showing efficacy of AI-based MR image reconstruction from undersampled acquisitions \cite{fastmri_results,modl_clinical,acc_factor_review_paper}.

For the fully sampled acquisition ($1\times$), no undersampling mask was applied.} Coil sensitivity maps were estimated using the ESPIRiT algorithm \cite{uecker2014espirit}. Complex Gaussian noise with a zero mean and a standard deviation of $15$ was added to the \acc{unnormalized} k-space data to emulate measurement noise at a level comparable to that observed in the fastMRI+ dataset.

\subsubsection{MRI Reconstruction Methods}
Two reconstruction methods were evaluated: a purely physics-based rSOS method and an AI-based U-Net method that performs denoising on rSOS-reconstructed images. Let $\bm{g}_i$ denote the k-space data acquired from the $i^{\text{th}}$ coil, and $\hat{f}_i$ represent its corresponding image obtained by applying the inverse Fourier transform,
\begin{equation}
\hat{f}_i = \mathcal{F}^{-1}\bm{g}_i.
\end{equation}
The rSOS reconstruction, denoted by $\hat{f}_{\text{rSOS}}$, was computed as
\begin{equation}
\hat{f}_{\text{rSOS}} = \sqrt{\sum_{i=1}^{N_c}|\hat{f}_i|^2},
\end{equation}
where $N_c = 8$ is the total number of coils.

The AI-based U-Net reconstruction network followed a classic encoder–decoder architecture. Both the encoder and decoder comprised four down-sampling and four up-sampling modules, respectively. Each module contained a convolutional block consisting of two consecutive $3 \times 3$ convolutions with a padding of 1 to preserve in-plane dimensions, each followed by batch normalization and ReLU activation. The encoder began with an initial convolutional block that mapped the single-channel input image to 64 feature maps, followed by three down-sampling modules that progressively increased the channel depth from 64 to 512 through max-pooling operations with a stride of 2, and a bottleneck layer expanding to 1024 channels. The decoder mirrored this structure, incorporating up-sampling modules that first upsampled feature maps by a factor of two using bilinear interpolation, applied a $1 \times 1$ convolution to adjust channel dimensionality. The feature maps were concatenated \infoclear{with} the corresponding encoder feature maps via skip connections. The combined features were then processed with a convolutional block. The number of feature channels decreased from 1024 to 64 across the decoder. A final $1 \times 1$ convolution produced a single-channel output with the same spatial dimensions as the input image. For each acceleration factor, the U-Net was trained to minimize the mean squared error between the rSOS-based accelerated reconstructions and fully sampled reference images using five-fold cross-validation on a dataset of 20,000 image pairs.

\begin{figure}[h]
\centering
\includegraphics[width=\linewidth]{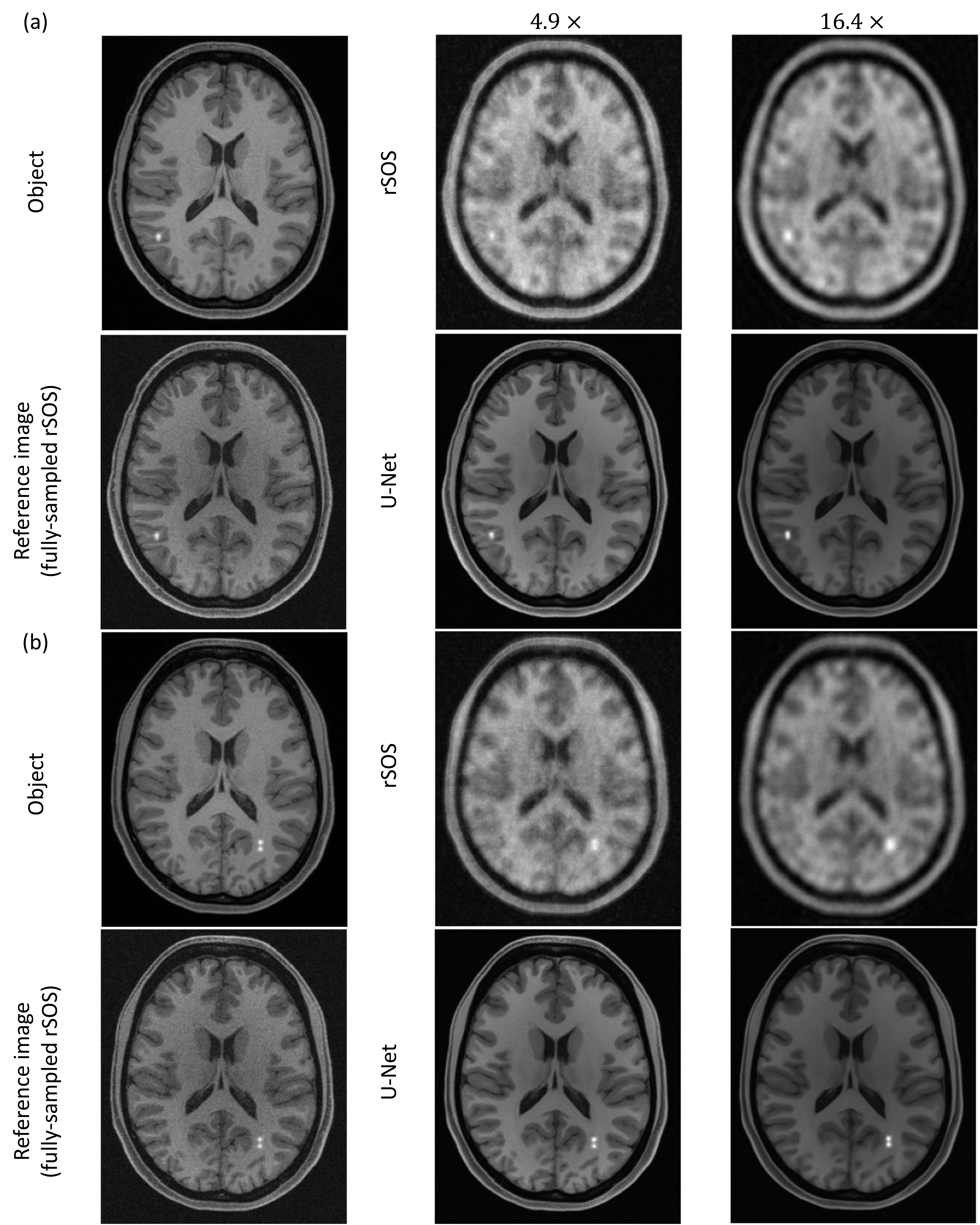}
\caption{Examples of objects and reconstructions, including (a) a singlet example and (b) a doublet example.}
\label{fig:obj_rec_examples}
\vspace{-1.0em}
\end{figure}

Examples of objects and reconstructions with singlet and doublet signals are shown in Fig.~\ref{fig:obj_rec_examples}.

\subsection{Training and Validation of DLMO}
\subsubsection{The DLMO Training} \label{sec:dlmo_train}
The DLMO was trained on a binary cross-entropy loss between the estimated probability of satisfying hypothesis $H_d$ and the corresponding ground truth labels, with optimization performed via the Adam algorithm \cite{kingma2014adam}. A transfer learning strategy was adopted to improve training efficiency and stability. Specifically, the network was first trained on images reconstructed from fully-sampled k-space data. Network weights were initialized using the Glorot normal scheme \cite{glorot2010understanding}, and all convolutional layers employed “same” padding to preserve spatial dimensions. To mitigate overfitting, dropout with a rate of 0.5 was applied after each LeakyReLU activation. The network was trained through five-fold cross-validation, and the final model was chosen based on validation performance, hereafter referred to as the base model. For each acceleration factor and reconstruction method, an additional DLMO was subsequently trained on accelerated MR reconstructions, with weights initialized from the base model, thereby refining the network to the specific acceleration factors and reconstruction methods. The number of training samples for both the base model and the refined DLMOs for each acceleration factor was 168,000. The training and validation were performed using PyTorch on four NVIDIA A100 GPUs, each with 80 GB of memory. Training curves of the base model training and refined model training are provided in the supplementary materials.

To ensure that the DLMO achieved performance comparable to human readers on the Rayleigh discrimination task, we adopted a human-label alignment training strategy. In this study, we considered singlet and doublet signals with different intensities and lengths, where signal length referred to the line length of singlets and the point separation of doublets. Instead of using all signal types in the training by assigning binary ground truth labels, images used in the DLMO training were selected based on human performance in a series of 2AFC studies. Each 2AFC study consisted of 50 image pairs reconstructed with the rSOS method and characterized by a specific combination of signal intensity, signal length, and acceleration factor. 
In each trial, a trained non-physician reader was asked to identify the image containing the doublet signal. For a given acceleration factor (i.e., either $1 \times$, $4.9 \times$ or $16.4 \times$), only those image sets with signal lengths and intensities where the reader achieved 100\% accuracy were included when training a DLMO for the considered acceleration factor. The included signal types are listed in the supplementary materials. Through this procedure, we obtained binary reference labels used for DLMO training, which were aligned with human decisions. Although the human reference labels were based on rSOS images, this procedure may also help ensure that any decay or gain in performance evaluated by a DLMO on an AI-based reconstruction aligns with that observed from human observers, which will be validated by a separate multi-reader multi-case study (described in the next section). In contrast, suppose a DLMO is trained purely on a ground truth signal set at the time of insertion into MR images as singlets and doublets, irrespective of signal size, separation length, contrast, or acceleration factor. This pure ground-truth–based idealized approach does not account for the limiting conditions of an MR imaging system, such as small-sized and minimal contrast-valued signals that can be humanly observed relative to noise levels and acceleration rates. As a result, such a procedure \infoclear{could} possibly yield a DLMO that performs at an exceedingly higher level than human observers. This DLMO \infoclear{could} exhibit high accuracy in detecting signals that humans may be unable to locate or discriminate. Consequently, when such an ideally-trained DLMO evaluates the gain or decay of an AI-based reconstruction, it becomes difficult to infer how that gain or decay \infoclear{could} equivalently manifest in the space of human perception. 

We further validated the performance of our human-label-trained DLMOs in a multi-reader multi-case human-observer study, with details provided in the next subsection.

\subsubsection{Validating human-label-trained DLMOs with Human Observers}
A multi-reader multi-case human observer study was conducted to validate the performance of the proposed DLMO against human readers. Prior research has demonstrated that 2AFC studies can be performed more efficiently than rating-based reader studies, with only a minor compromise in variance \cite{burgess1995comparison, shekter2020efficiently}. Theoretical analyses further indicate that the proportion correct achieved by an observer in a 2AFC study is equivalent to the area under the receiver operating characteristic (ROC) curve (AUC) obtained in a corresponding rating-based study \cite{barrett2013foundations, liu2023observer}. Based on these findings, in this study, human performance on the Rayleigh discrimination task was assessed in 2AFC studies. The objective of this validation was to determine whether human-label-trained DLMO's performance was statistically similar to that of human readers within a pre-specified similarity margin of 0.1 proportion correct. \acc{Because the objective of this proof-of-concept study was to develop a DLMO whose performance trends are consistent with those of human readers, rather than to establish equivalence between reconstruction methods, we adopted a relatively relaxed non-inferiority margin of 0.1. This choice is supported by the substantial variability in absolute AUC performance observed across human readers in published MRMC studies, such as Conant et al. \cite{conant2019improving}, in which reader AUCs ranged from 0.631 to 0.905.}

The signal intensities and lengths used for this validation study were selected for each acceleration factor to ensure an appropriate task difficulty. Specifically, the parameters were chosen such that the task was neither too challenging for human readers to provide meaningful assessments nor too trivial to result in near-perfect accuracy, based on the observation in the human-label alignment training strategy mentioned in \ref{sec:dlmo_train}. The corresponding signal characteristics are summarized in Table~\ref{tab:signal_char}.

\begin{table}[h]
    \caption{Intensities and lengths of signals in the test dataset.}
    \begin{center}
    \label{tab:signal_char}
    \begin{NiceTabular}{c|c|c}
	\hline
	\textbf{Acceleration factor} & \textbf{Signal intensity} & \textbf{Signal length} (mm) \\
	\hline
	 $4.9 \times$ &  0.7 & 5, 6, 7, and 8\\
    \hline
	 $16.4 \times$ &  1.3 & 5, 6, 7, and 8\\
    \hline
	\end{NiceTabular}
    \end{center}
\end{table}

We performed a sample size calculation to determine the sample size for this validation. A pilot study involving three trained non-physician readers and 50 pairs of singlet and doublet images was first conducted. Also, the performance of the DLMO was estimated with 4,000 pairs of images. Our sample size calculation revealed that for the considered similarity margin, to show the similarity of DLMOs to human readers with a significance level of 0.05, $N_s=640$ singlet and $N_d=640$ doublet samples were \blc{warranted} to be read by $N_R=4$ readers in a split-plot study. Details of the sample size calculation are provided in the supplementary materials.

Accordingly, the pivotal study included $640$ singlet–doublet pairs and four trained non-physician readers. Each of the four readers separately evaluated $160$ independent image pairs in a 2AFC study (i.e., non-overlapping cases between the readers). Each reader evaluated their corresponding 160 independent image pairs for two reconstruction methods (rSOS and U-Net) and two acceleration factors ($4.9 \times$ and $16.4 \times$). Thus, each reader completed four 2AFC studies, evaluating a total of 640 image pairs ($2 \times 2 \times 160 = 640$).
Each image pair contains singlet and doublet images with the same signal intensity and length and was assigned to exactly one reader. The reader studies were conducted using a web-based application \cite{liu2023observer}. The proportion correct was computed for both rSOS and U-Net reconstructions in each 2AFC study, and the differences in proportion correct between the DLMO and human readers, along with their corresponding variances, were subsequently estimated using the iMRMC application \cite{imrmc}. Since readers interpreted images reconstructed by two reconstruction methods at two acceleration factors, Bonferroni correction\footnote{The Bonferroni correction adjusts probability (p) values because of the increased risk of a type I error when making multiple statistical tests \cite{armstrong2014use}.} was applied when testing for similarity between DLMO and human observers.

\subsection{Evaluations}
\subsubsection{Extract Task-specific Information}
The performance of rSOS and U-Net reconstructions were evaluated on the Rayleigh discrimination task using the proposed DLMO. In this task, the background was represented by a random vector drawn from a non-degenerate distribution, and the signals were also random, corresponding to a signal-known-statistically and background-known-statistically (SKS/BKS) detection paradigm\cite{barrett2013foundations,zhou2019approximating}. Signal intensity and length were the same as those listed in Table~\ref{tab:signal_char}, with the signal length extended to include 4 mm.

The test dataset comprised 4,000 singlet and 4,000 doublet samples. Reconstructed images from both the rSOS and U-Net methods were normalized to the range from 0 to 1 prior to being input into the DLMO. The scalar output of the DLMO prior to the sigmoid activation function was considered as the test statistic, which was compared to a threshold to classify each test sample as either with a singlet or a doublet. By varying this threshold, we calculated true-positive and false-positive rates and plotted ROC curves.

\subsubsection{Figures of Merit}
We calculated AUC values with 95\% confidence intervals (CIs) for rSOS and U-Net reconstruction methods and differences in AUCs with 95\% CIs between these methods, across different signal lengths and acceleration factors. To account for multiple hypothesis testing, Bonferroni correction was applied.

Further, to assess the visual similarity between accelerated and fully-sampled MR images, we computed the PSNR and SSIM between accelerated MR reconstructions and fully-sampled MR reconstructions and compared these values using \textit{t}-test.

For all statistical tests in this study, a p-value less than 0.05 was used to infer statistical significance.

\section{Results}
\subsection{Validation of DLMO with Human Observers}
\label{sec:2afc_validation}
Table~\ref{tab:2afc_validation} summarizes the difference in proportion correct between DLMO and human observers with corresponding 95\% CIs. We observed that the CIs of difference in proportion correct between DLMO and human observers were within a margin of the pre-specified similarity margin of 0.1 proportion correct, across reconstruction methods and acceleration factors. Thus, the performance of DLMO was deemed to be statistically similar to human observers within the pre-specified margin \cite{ahn2013demonstrate}. Furthermore, Table~\ref{tab:2afc_validation_shorter_length} presents the mean differences in proportion correct between the DLMO and human observers for images with shorter signal lengths (5 and 6 mm), including those were not used during DLMO training, and longer signal lengths (7 and 8 mm). Across both acceleration factors and reconstruction methods, the mean differences in proportion correct between the DLMO and human observers were all below 0.1. The alignment between DLMO and reader performance supports its reliability as a surrogate for human observers in this study.

\begin{table*}[h]
    \caption{Difference in proportion correct between DLMO and human observers}
    \begin{center}
    \label{tab:2afc_validation}
    \begin{NiceTabular}{c|c|c|c|c|c|c}
	\hline
	\textbf{Acceleration} &
\multicolumn{3}{c}{\multirow{2}{*}{\textbf{rSOS (CIs)}}} &
\multicolumn{3}{c}{\multirow{2}{*}{\textbf{U-Net (CIs)}}}\\
\textbf{factor} &\multicolumn{3}{c}{ } &\multicolumn{3}{c}{ } \\
	\hline
	  \multirow{2}{*}{$4.9 \times$} & Human observers & DLMO & $\Delta$ & Human observers & DLMO & $\Delta$ \\
       & 0.91 (0.86, 0.95) & 0.95 (0.95, 0.95) &  0.043 (-0.0080, 0.095) & 0.95 (0.92, 0.98) & 0.98 (0.98, 0.98) & 0.029 (-0.0082, 0.066)\\
    \hline
	  \multirow{2}{*}{$16.4 \times$} & Human observers & DLMO & $\Delta$ & Human observers & DLMO & $\Delta$ \\
       & 0.89 (0.85, 0.93) &  0.95 (0.95, 0.96) & 0.065 (0.034, 0.095) & 0.96 (0.93, 0.98) & 0.96 (0.96, 0.96) & 0.0026 (-0.019, 0.024)\\
    \hline
    \multicolumn{7}{l}{CIs: 95\% confidence intervals.} \\
    \multicolumn{7}{l}{$\Delta$: Difference between human observers and DLMO.} \\
    \multicolumn{7}{l}{Signal intensities are 0.7 and 1.3 at acceleration \infoclear{factors} of 4.9 and 16.4, respectively.}
	\end{NiceTabular}
    \end{center}
\end{table*}

\begin{table*}[h]
    \caption{Mean difference in proportion correct between DLMO and human observers with different signal lengths}
    \begin{center}
    \label{tab:2afc_validation_shorter_length}
    \begin{NiceTabular}{c|c|c|c|c|c|c|c}
	\hline
	\textbf{Signal} & \textbf{Acceleration} &
\multicolumn{3}{c}{\multirow{2}{*}{\textbf{rSOS}}} &
\multicolumn{3}{c}{\multirow{2}{*}{\textbf{U-Net}}}\\
\textbf{length} & \textbf{factor} &\multicolumn{3}{c}{ } &\multicolumn{3}{c}{ } \\
	\hline
	 \multirow{4}{*}{5 and 6 mm}  & \multirow{2}{*}{$4.9 \times$} & Human observers & DLMO & $\Delta$ & Human observers & DLMO & $\Delta$ \\
      & & 0.85 & 0.91 &  0.057 & 0.92 & 0.97 & 0.048\\
    \cline{2-8}
	  & \multirow{2}{*}{$16.4 \times$} & Human observers & DLMO & $\Delta$ & Human observers & DLMO & $\Delta$ \\
      & & 0.83 &  0.91 & 0.087 & 0.93 & 0.94 & 0.0035\\
    \hline
	  \multirow{4}{*}{7 and 8 mm}  & \multirow{2}{*}{$4.9 \times$} & Human observers & DLMO & $\Delta$ & Human observers & DLMO & $\Delta$ \\
      & & 0.97 & 0.99 &  0.030 & 0.98 & 0.99 & 0.010\\
    \cline{2-8}
	  & \multirow{2}{*}{$16.4 \times$} & Human observers & DLMO & $\Delta$ & Human observers & DLMO & $\Delta$ \\
      & & 0.95 &  0.99 & 0.043 & 0.98 & 0.99 & 0.0018\\
    \hline
    \multicolumn{8}{l}{$\Delta$: Difference between human observers and DLMO.} \\
    \multicolumn{8}{l}{Signal intensities are 0.7 and 1.3 at acceleration \infoclear{factors} of 4.9 and 16.4, respectively.}
	\end{NiceTabular}
    \end{center}
\end{table*}

\subsection{Evaluation Based on Fidelity-Based Figures of Merit}
\label{sec:fidelity}
Table~\ref{tab:psnr_ssim} shows PSNR and SSIM values obtained for the rSOS and U-Net reconstructions at acceleration factors of $4.9 \times$ and $16.4 \times$ with rSOS at $1 \times$ (i.e., fully sampled reconstruction) set as reference images. Across both acceleration factors, U-Net achieved markedly higher PSNR and SSIM values compared with rSOS, indicating improved noise suppression and structural preservation in the reconstructed images. These results demonstrate that the U-Net reconstruction produced visually smoother, more pleasing and more structurally coherent images than rSOS.

\begin{table}[h]
    \caption{PSNR and SSIM values obtained using rSOS and U-Net with 4,000 singlet and 4,000 doublet samples}
    \begin{center}
    \label{tab:psnr_ssim}
    \begin{NiceTabular}{c|c|c|c|c}
	\hline
	\textbf{\shortstack{Acceleration\\factor}} & \multicolumn{2}{c}{\textbf{PSNR} (std)} & \multicolumn{2}{c}{\textbf{SSIM} (std)} \\
	\hline
	  \multirow{2}{*}{$4.9 \times$}
       &  rSOS & 28.39 (0.50) &  rSOS & 0.6777 (0.01)\\
       &  U-Net & \textbf{33.94 (0.36)} &  U-Net & \textbf{0.8183 (0.01)}\\
    \hline
	  \multirow{2}{*}{$16.4 \times$}
       &  rSOS & 26.64 (0.51) &  rSOS & 0.5941 (0.01)\\
       &  U-Net & \textbf{33.54 (0.55)} &  U-Net & \textbf{0.8080 (0.01)}\\
    \hline
	\end{NiceTabular}
    \end{center}
\end{table}

\subsection{Evaluation on the Rayleigh Discrimination Task}

\begin{figure}[h]
\centering
\includegraphics[width=0.8\linewidth]{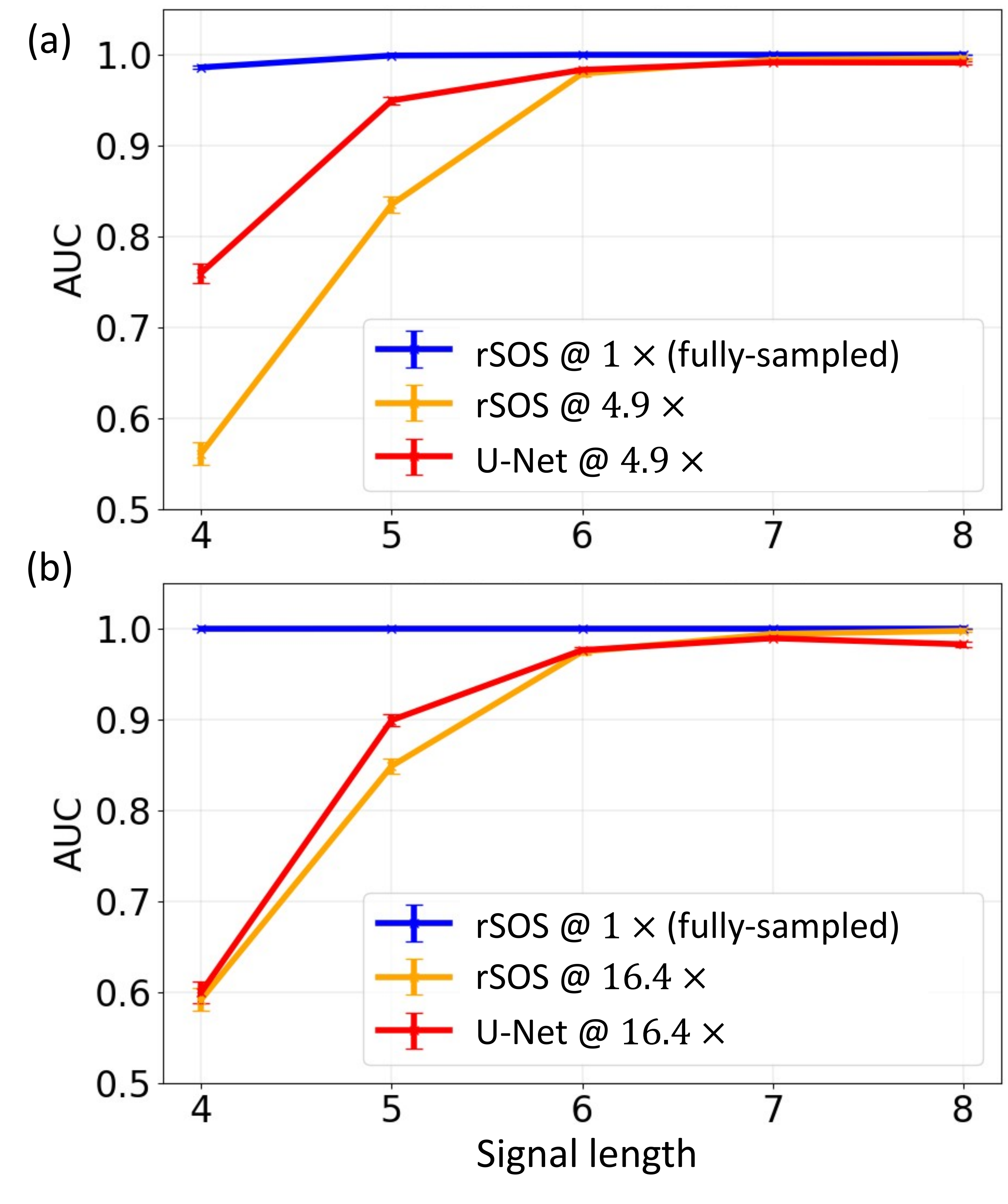}
\caption{AUC values obtained by DLMO with rSOS and U-Net reconstructions at acceleration factors of (a) $4.9 \times$ and (b) $16.4 \times$, with 4,000 singlet and 4,000 doublet samples. Note that signal intensity was 0.7 at \infoclear{an} acceleration factor of $4.9 \times$ and 1.3 at $16.4 \times$.}
\label{fig:AUC_unet_rsos}
\vspace{-1.0em}
\end{figure}

Fig.~\ref{fig:AUC_unet_rsos} shows the resolution performance of the rSOS and U-Net reconstruction approaches on the Rayleigh discrimination task, as evaluated using the proposed DLMO. We observed that increasing the acceleration factor resulted in a degradation of image resolution for both reconstruction approaches. At an acceleration factor of $4.9 \times$, the U-Net exhibited significantly higher performance than rSOS for shorter signal lengths (4 mm and 5 mm), while showing similar performance for longer signals (6–8 mm). The performance of the DLMO in discriminating shorter-length signals (i.e., $< 6$ mm) at $4.9 \times$ aligns with the conclusion drawn from the fidelity-based analysis (i.e., the performance of \infoclear{the} U-Net at $4.9 \times$ is better than that of rSOS at $4.9 \times$). However, the DLMO-based Rayleigh discrimination task provides important information not captured by PSNR or SSIM. Specifically, although U-Net at $4.9 \times$ performs better than rSOS at $4.9 \times$ for shorter signals ($< 6$ mm), the AUC for U-Net $4.9 \times$ was substantially lower than that of rSOS at $1 \times$, which represents the standard-of-care method.

At $16.4 \times$ acceleration, however, U-Net performance was similar to or slightly inferior to rSOS for most signal lengths, except for the 5 mm case. Notably, at this high acceleration factor, the conclusions drawn from fidelity-based FoMs and those derived from task-based evaluation were simply inconsistent. These observations highlight that, although improvements in PSNR or SSIM for linear reconstruction methods often align with gains in diagnostic accuracy, this relationship does not necessarily extend to non-linear, AI-based models such as the U-Net used in our study. This phenomenon was consistent with previous reports of deep learning methods producing visually smoother images with reduced apparent noise but with limited task-based performance \cite{yu2020ai,yu2023need,li2021assessing,kc2024assessing}.

\subsection{Visual Inspection} \label{sec:visual}
\begin{figure}[h]
\centering
\includegraphics[width=\linewidth]{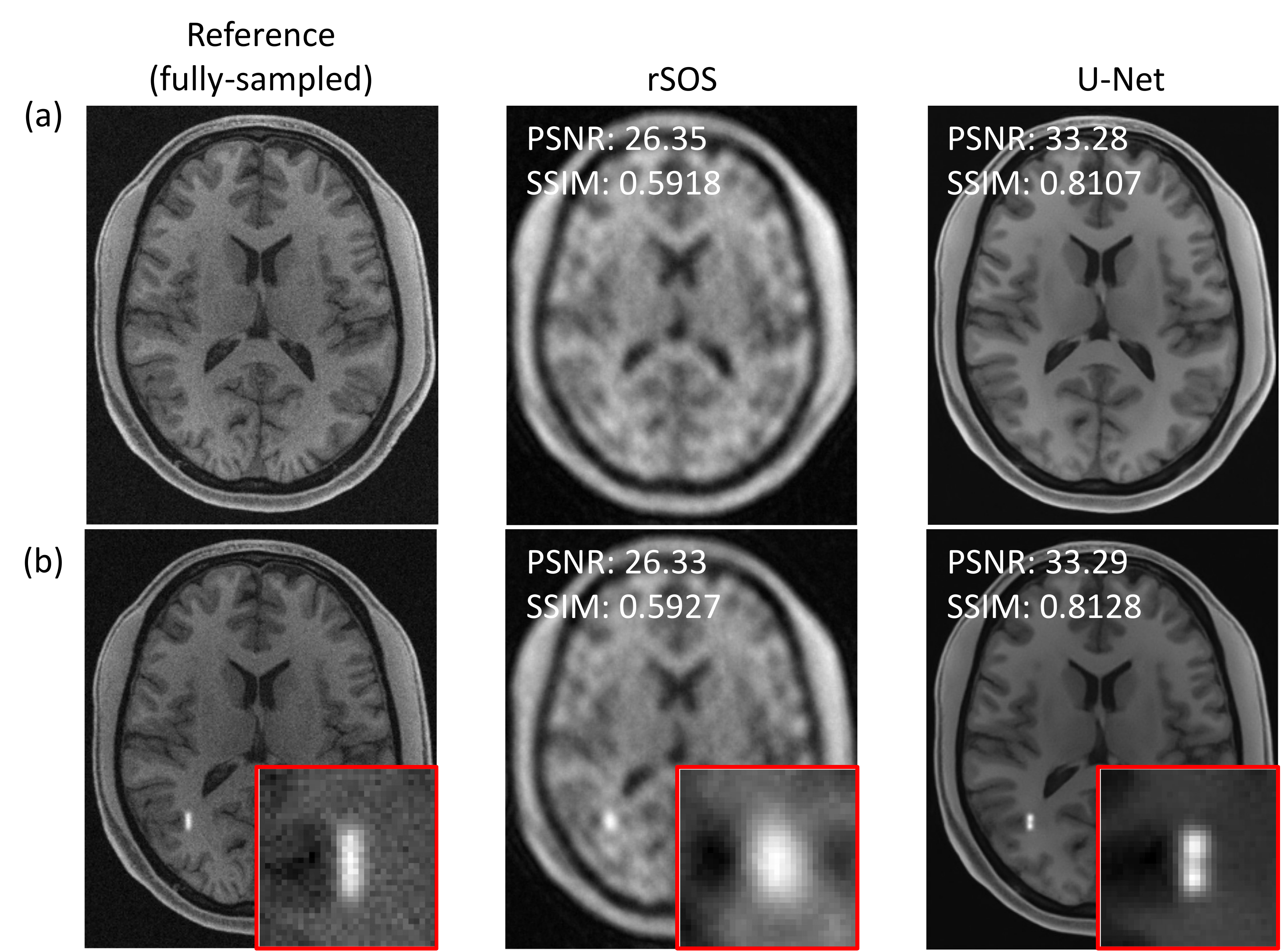}
\caption{Representative reconstructions at an acceleration factor of $16.4 \times$. (a) rSOS ($1 \times$), rSOS ($16.4 \times$), and U-Net ($16.4 \times$) reconstructions without any inserted signals. (b) singlet-signal examples reconstructed under the same conditions. In (b), the U-Net reconstruction hallucinated the singlet signal, producing an appearance resembling a doublet signal.}
\label{fig:rec_examples}
\vspace{-1.0em}
\end{figure}

Fig.~\ref{fig:rec_examples} serves as a representative example of images reconstructed using the rSOS and U-Net methods. The first row compares reference rSOS ($1 \times$), accelerated rSOS ($16.4 \times$), and accelerated U-Net ($16.4 \times$) reconstructions without any inserted signals. Based on visual appearance, the U-Net reconstruction appeared smoother and more similar to the fully sampled reference image than rSOS ($16.4 \times$), consistent with its superior PSNR and SSIM values. Indeed, one might even conclude from visual inspection that U-Net ($16.4 \times$) performs comparably to rSOS ($1 \times$), suggesting minimal degradation at high acceleration factors.

However, this perceived superiority did not hold under task-based evaluation. The second row of Fig.~\ref{fig:rec_examples} shows singlet-signal examples reconstructed under the same conditions. Here, the U-Net ($16.4 \times$) reconstruction introduced a hallucinated doublet-like feature that was absent in both the reference and rSOS images. Yet, the U-Net reconstruction yielded superior PSNR and SSIM values than rSOS ($16.4 \times$). This subtle but clinically meaningful error demonstrates that AI-based reconstructions \infoclear{may} generate plausible yet incorrect structural content. Such artifacts \infoclear{may} easily go unnoticed in routine side-by-side visual inspection or when using conventional fidelity-based metrics. In this case, a likely explanation is that the deep-learning model attempts to compensate for undersampled k-space data by enforcing spatial symmetry or smoothness, inadvertently converting a true singlet into a doublet-like structure.

These observations show that visual inspection and fidelity-based FoMs (e.g., PSNR and SSIM) are agnostic to diagnostic performance. In contrast, the proposed \infoclear{DLMO approach} directly quantifies image quality on a predefined clinical task. This task-based evaluation captures clinically relevant degradation that purely appearance-based or fidelity-based FoMs cannot.


\section{Discussion}
In this work, we developed and validated a deep learning–based model observer (DLMO) for task-based evaluation of accelerated MR reconstruction methods. Using the proposed \infoclear{DLMO} within an in-silico imaging trial, we assessed a conventional physics-based rSOS reconstruction and an AI-based U-Net reconstruction on the Rayleigh discrimination task, \infoclear{which acts} as a surrogate measure of resolution. 

The ultimate goal of medical image reconstruction is to convert raw measurement to images that are interpretable by human. Thus, we calibrated the performance of the DLMO to human readers when evaluating the performance of image reconstruction or denoising methods. The proposed DLMO showed similarity with human readers within a pre-specified margin of 0.1 proportion correct in 2AFC studies (Table~\ref{tab:2afc_validation} and Table~\ref{tab:2afc_validation_shorter_length}), demonstrating its ability to capture task-relevant image features that influence perceptual ability to discriminate. The use of human-label alignment in training played a key role in this outcome. By calibrating the DLMO using reader responses from cases where humans achieved perfect accuracy, we ensured that the model learned features that were both task-specific and perceptually meaningful, rather than being tied to idealized or algorithmic labels. The incorporation of multiple signal boundary conditions for each given acceleration (shown in Fig.3 in the supplementary materials) enabled the DLMO to learn a finely graded resolution with respect to both increases and decreases in signal length and signal intensity. This is substantiated by the subgroup analysis for shorter (5 and 6 mm) and longer (7 and 8 mm) signal as shown in Table~\ref{tab:2afc_validation_shorter_length}. Furthermore, the transfer learning strategy allowed the DLMO to adapt efficiently to varying acceleration factors and reconstruction methods without the \infoclear{necessity} for retraining from scratch, reducing computational demands while maintaining robustness. Thus, the proposed \infoclear{DLMO approach may} be reapplied to other image acquisition settings and reconstruction methods.

The image fidelity-based evaluation in section \ref{sec:fidelity} demonstrated that U-Net at $4.9 \times$ and $16.4 \times$ achieved significantly higher PSNR and SSIM values than rSOS at the same acceleration factors. However, these metrics failed to capture the subpar performance of U-Net in a discrimination task compared to the clinically acceptable baseline of rSOS at $1 \times$ (i.e., fully sampled reconstruction). Specifically, using the proposed \infoclear{DLMO approach} on the Rayleigh discrimination task, we observed that U-Net at $4.9 \times$ outperformed rSOS at $4.9 \times$ for short-length signals ($4–5$ mm). Yet, rSOS at $1 \times$ outperformed U-Net at $4.9 \times$ for shorter signals (i.e., U-Net’s AUC dropped by approximately 25\% for $4$ mm and 5\% for $5$ mm signals). Further increasing the acceleration factor revealed a corresponding decline in U-Net’s performance. In particular, U-Net at $16.4 \times$ performed similarly to or worse than rSOS at $16.4 \times$. Most importantly, the performance of U-Net at $16.4 \times$ was substantially poorer than that obtained using rSOS at $1 \times$. For instance, the AUC of U-Net at $16.4 \times$ dropped by about 40\% for $4$ mm and 10\% for $5$ mm signals those from rSOS at $1 \times$.

In section \ref{sec:visual}, we noted that visual inspection, similar to image fidelity–based analysis, offers an assessment of perceptual image quality (i.e., how visually pleasing an image appears when compared side by side). We elaborated that this type of evaluation is inherently subjective, lacks an explicitly defined clinical task within its \infoclear{analysis}, and may not accurately represent the clinical reliability of a reconstruction.

The findings from image fidelity–, visual inspection–, and DLMO-based analyses indicate that although AI-based accelerated reconstructions may produce smoother and more visually appealing images, they can compromise task-relevant resolution as the acceleration rate is sequentially increased (or as the acquisition time is shortened). Therefore, a task-based evaluation method (such as our DLMO \infoclear{approach} on the Rayleigh discrimination task) \infoclear{may be} useful to objectively evaluate and compare the performance of new reconstruction methods against \infoclear{existing method} in preserving diagnostically important information.

Our study also demonstrates the feasibility of integrating in-silico imaging trials with model observers to perform scalable, reproducible evaluations of image reconstruction methods. The proposed DLMO \infoclear{approach} allows systematic variation of task conditions, such as signal contrast and size, and spatial location, across a large number of simulated patients without the logistical and ethical constraints of clinical trials. Within this in-silico environment, the brain anatomy background can be generated from stochastic object models that emulate patient variability. These clinically relevant brain background could be also generated from other learning-based generative models or virtual-patient-generating software \cite{segars20104d,badal2019virtual}. Based on in-silico imaging trials, the proposed \infoclear{DLMO} can be employed to find feasible acceleration factors for brain MR or other radiological modalities. Through the combination of statistical object models, simulated imaging systems, and computational observers, the proposed \infoclear{DLMO approach} offers a \infoclear{preclinical} cost-effective \infoclear{technique} for evaluating MRI reconstruction methods.

There are several limitations to this work. First, the current study only focused on the Rayleigh discrimination task as a surrogate for resolution of reconstruction methods. However, given the promising performance observed in this study, the proposed DLMO could be extended to other clinically relevant tasks, including lesion detection \cite{li2025estimating} and quantification. To assess other issues, such as hallucinations \cite{granstedt2025hallucinations}  observed in Fig.~\ref{fig:rec_examples}, \blc{a different testing approach may be appropriate} \cite{kc2026sfrc,bhadra2021hallucinations}. Second, in this study, we used a U-Net–based post-processing network as a representative AI-based reconstruction method due to its simplicity and widespread use in accelerated MRI. However, the proposed \infoclear{DLMO approach} itself is not limited to U-Net. The same \infoclear{approach may} be extended to other deep learning reconstruction methods, such as unrolled optimization methods or physics-integrated neural networks. Future work may explore the assessment of feasible acceleration factors with these more advanced reconstruction approaches. Third, although synthetic brain phantoms generated by the DDPM provided realistic appearance of background structures and were validated to have similar distributions of gray and white matter ratios to those of real brain MR images, some of DDPM-generated images contained artifacts \cite{deshpande2025rep}. During our data inspection process, some of the DDPM-generated brain images were found to contain obvious skull artifacts (cut-offs or non-smooth skull borderlines), and it was impossible to manually identify all of those artifacts in a large dataset. Despite the possible presence of artifacts at the skull of the brain images created by the generative model, the overall structure still provided much more realistic anatomical background statistics than uniform background for testing novel AI-enabled MRI image reconstruction methods. In the future, automatic detection and removal of those artifacts-containing synthetic images may be developed to minimize the impact of these artifacts on the estimated image performance. It is also important to note that this study does not investigate issues related to generalizability, subgroup analysis, or patient-/acquisition-based covariate effects. Rather, before addressing these important considerations, this work questions whether widely used PSNR/SSIM and Likert-based analyses remain appropriate for evaluating the clinical efficacy of new reconstruction methods in the era of AI.

In summary, the proposed \infoclear{DLMO} provides an efficient, objective, and human-aligned approach for evaluating AI-based accelerated MR reconstruction methods. While the considered U-Net reconstruction method produced smoother and more visually appealing reconstructions, evaluation on the Rayleigh discrimination task revealed a potential trade-off between apparent image quality and preservation of diagnostically relevant detail. In addition, the \infoclear{DLMO} enables objective, task-based assessment of reconstruction quality \infoclear{while avoiding} physical data acquisition or large-scale human reader studies. This makes it a scalable and minimally burdensome tool for evaluating emerging reconstruction methods. Moreover, the procedure described in this \infoclear{contribution} can be readily implemented by AI researchers within a computational environment to explore the loss of diagnostic task information due to MR acceleration during algorithm development.

\section{Conclusion}

This study presented a deep learning–based model observer \infoclear{(DLMO)} for evaluation of accelerated MR reconstruction methods on the Rayleigh discrimination task, demonstrated through an in-silico imaging trial comparing a conventional rSOS approach with an AI-based U-Net reconstruction. The DLMO was trained to calibrate with human-observer performance and validated in a multi-reader two-alternative forced choice  study, effectively capturing human-like discrimination performance while remaining computationally efficient through transfer learning. While the U-Net consistently improved fidelity and reduced noise, evaluation on the Rayleigh discrimination task using the DLMO revealed similar or inferior performance of U-Net to rSOS with fully-sampled data.
These results emphasize the importance of integrating task-based assessments into the evaluation of AI-based reconstruction algorithms. The proposed \infoclear{DLMO} contributes a scalable, reproducible, and human-aligned alternative to real patient data and real acquisitions for exploring diagnostically meaningful MR acceleration rates. It is less burdensome and avoids labor-intensive procedures and costly real human data collection.

\vspace{5mm}
\textbf{Code and data}
All the codes and data employed in the numerical studies can be found at \url {https://github.com/DIDSR/dlmo}.

\textbf{Disclaimer}
This article reflects the views of the authors and does not represent the views or policy of the U.S. Food and Drug Administration, the Department of Health and Human Services, or the U.S. Government.  The mention of commercial products, their sources, or their use in connection with material reported herein is not to be construed as either an actual or implied endorsement of such products by the Department of Health and Human Services.

\textbf{Acknowledgment} We thank Drs. Kaiyan Li, Mark Anastasio and Hua Li for sharing the code of generating synthetic brain data with trained generative models that was developed in \cite{li2025estimating}. We also thank Dr. Brandon Gallas for his guidance on the statistical analyses and Dr. Andrey Makeev for discussions on the DLMO training strategy. Zitong Yu acknowledges funding by appointment to the research participant program at the Center for Devices and Radiological Health administrated by the Oak Ridge Institute for Science and Education through an interagency agreement between the US Department of Energy and the US Food and Drug Administration.

\bibliographystyle{IEEEtran}
\bibliography{IEEEabrv}

\end{document}


\title{Evaluating the resolution of AI-based accelerated MR reconstruction using a deep learning-based model observer: Supplementary materials}

\maketitle

\section{Synthetic brain data generation using DDPM}
We used a denoising diffusion probabilistic model (DDPM) to generate synthetic brain data \cite{ho2020denoising}. The DDPM was previously trained by Li et al. \cite{li2025estimating} and is available in our GitHub repository \cite{dlmo_github}. Table~\ref{tab:stat_ddpm} summarizes key image statistics, including image mean, image standard deviation, gray and white matter ratios, estimated from both the DDPM-generated images and the Human Connectome Project (HCP)'s \cite{van2013wu,HCP_dataset} real brain MR images. These statistics were highly similar between the two datasets.
\begin{table*}[h]
    \caption{Statistical and area-based comparisons between $1,000$ real clinical images and $1,000$ DDPM-generated synthetic images.}
    \begin{center}
    \label{tab:stat_ddpm}
    \begin{NiceTabular}{c|c|c|c|c}
	\hline
	& \textbf{Mean} & \textbf{STD} & \textbf{$R_{white}$} & \textbf{$R_{gray}$} \\
	\hline
    Real Clinical Images & $0.30\pm 0.035$ & $0.23\pm 0.020$  & $0.31\pm 0.11$ & $0.42\pm 0.082$ \\
	\hline
    DDPM Synthetic Images & $0.33\pm 0.057$ & $0.25\pm 0.032$  & $0.33\pm 0.093$ & $0.39\pm 0.069$\\
    \hline
    \multicolumn{5}{l}{STD: Standard deviation.}\\
    \multicolumn{5}{l}{$R_{white}$: the ratio of white-matter area to the area inside the skull.}\\
    \multicolumn{5}{l}{$R_{gray}$: the ratio of gray-matter area to the area inside the skull.}
	\end{NiceTabular}
    \end{center}
\end{table*}

\begin{figure}[h]
\centering
\includegraphics[width=0.8\linewidth]{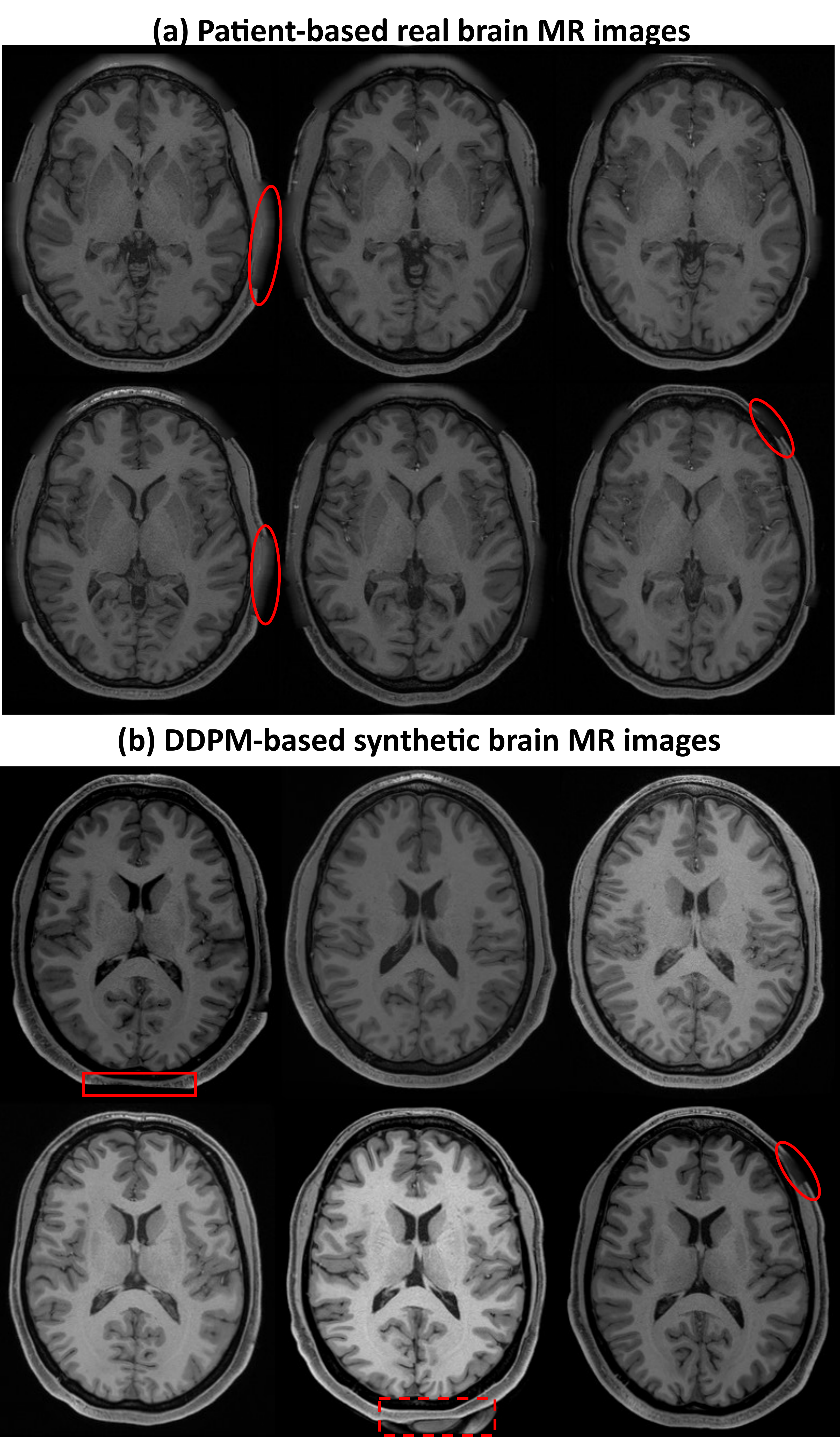}
\caption{Examples of (a) real images from the Human Connectome Project (HCP) brain MR dataset and (b) DDPM-generated brain MR images. Red ovals in both real and denoising diffusion probabilistic model (DDPM)-generated images indicate commonly observed skull artifacts caused by metal implants or bobby pins. The red rectangular box in the top-left corner of plot (b) depicts a field-of-view (FOV) cut-off artifact. The dotted rectangular box in the bottom center of plot (b) illustrates a wrap-around artifact combined with wrapping of anatomy outside the skull. Overall, DDPM-generated images exhibit skull artifacts observed in real images, as well as new forms of skull artifacts.}
\label{fig:hallu_ddpm}
\end{figure}

\begin{figure*}[!tbh]
\centering
\includegraphics[width=0.7\linewidth]{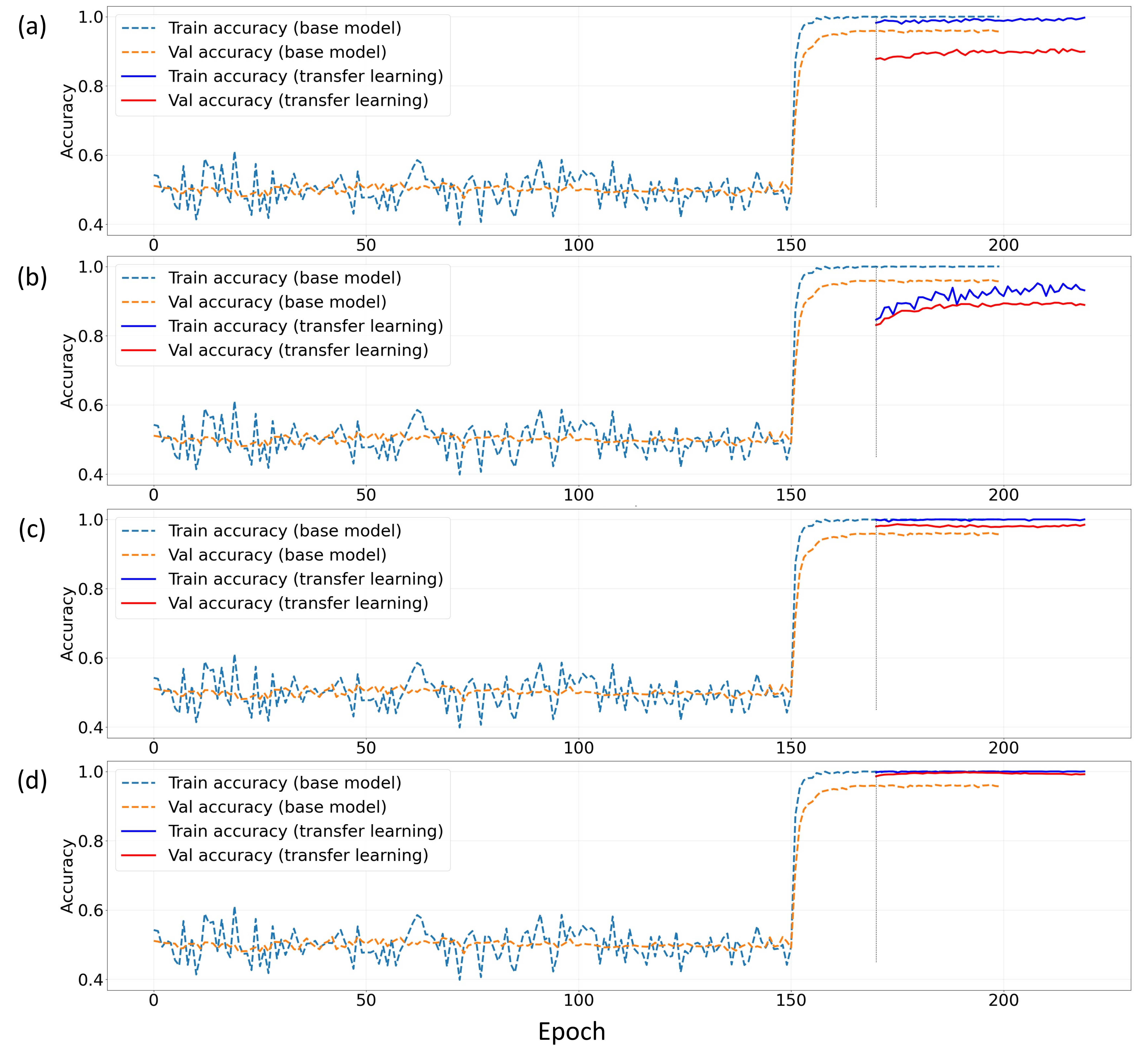}
\caption{Convergence curves of the base model training (dashed lines) and refined model training (solid lines). (a) and (b) are curves for DLMOs trained on rSOS reconstructed images at $4.9 \times$ and $16.4 \times$, respectively, while (c) and (d) are curves for DLMOs trained on U-Net reconstructed images at $4.9 \times$ and $16.4 \times$, respectively. Transfer learned models were trained on base models at epoch of 170.}
\label{fig:training_curve}
\end{figure*}

As mentioned in the Discussion section in the main paper, some DDPM-generated images exhibited skull-related artifacts, such as cut-offs or irregular skull boundaries (Fig.~\ref{fig:hallu_ddpm}). However, because singlet and doublet signals were inserted only within white or gray matter regions, and white-gray matter distribution is statistically similar to those observed in real images. Thus, these skull artifacts were unlikely to affect the task-based evaluation.

\section{Convergence curves of the base model training and refined model training}
Fig.~\ref{fig:training_curve} shows the convergence curves for the base deep learning-based model observer (DLMO) model (trained on fully sampled $1 \times$ data) and for the refined models obtained through transfer learning at $4.9 \times$ and $16.4 \times$ acceleration. The refined models were initialized from the base model trained on rSOS reconstructed images at epoch 170. The transfer-learned models converged faster than the base model while achieving comparable discrimination accuracy. This demonstrates that the transfer-learning strategy enables the DLMO to efficiently adapt to different acceleration factors and reconstruction methods without requiring full retraining, thereby reducing computational burden while preserving model robustness.

\section{Determination of signal parameters for the DLMO training dataset}

\begin{figure}[h]
\centering
\includegraphics[width=\linewidth]{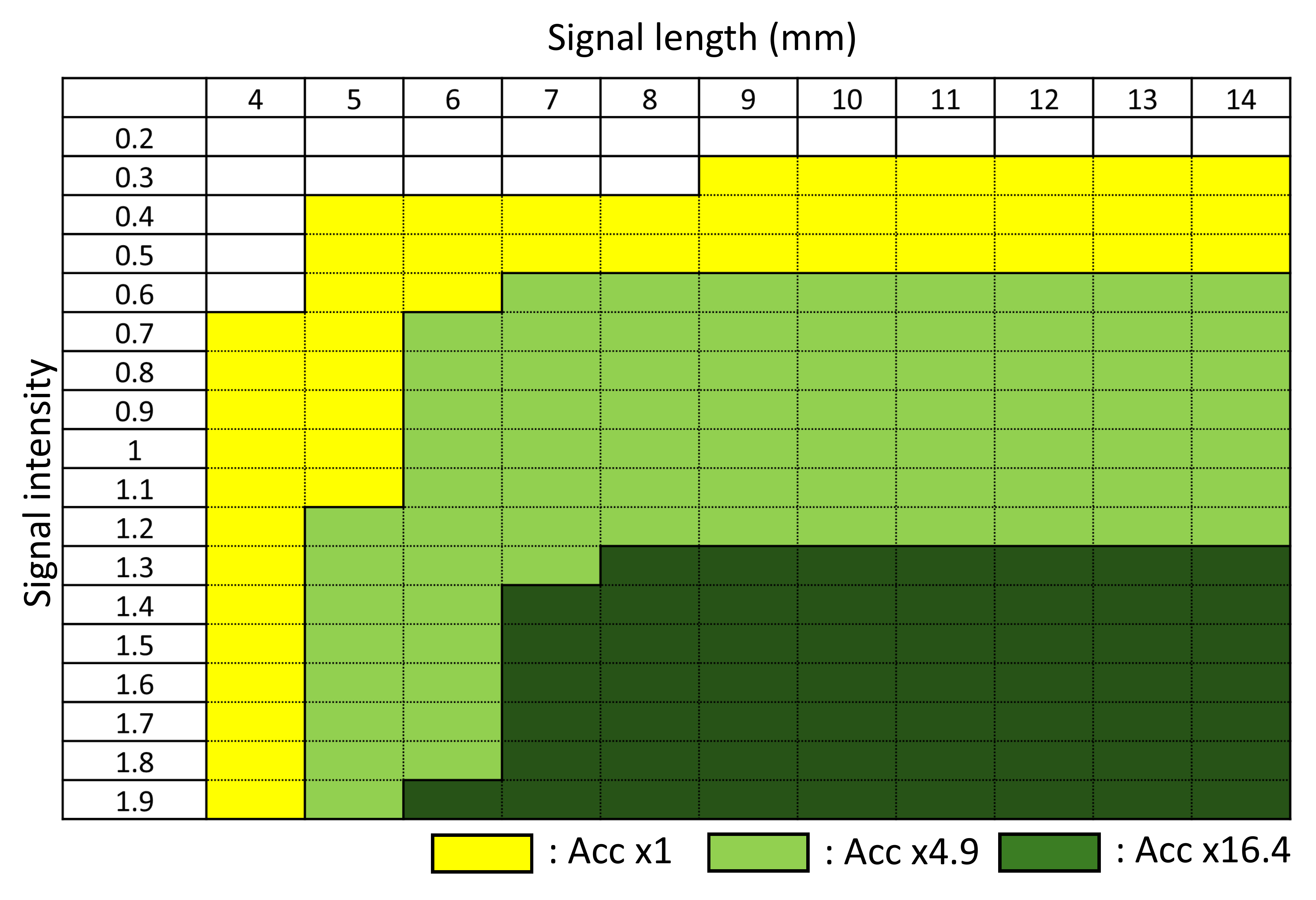}
\caption{Performance of the trained non-physician reader across a series of 2AFC studies. Each cell corresponds to one 2AFC study defined by a specific combination of signal intensity and signal length. Colored cells indicate conditions under which the reader achieved 100\% accuracy on the Rayleigh discrimination task. Note that whenever 100\% accuracy was achieved at a higher acceleration factor, the same signal condition also yielded 100\% accuracy at lower acceleration factors. For example, a signal with intensity 1.3 and length 8 mm resulted in perfect accuracy at acceleration factors of $16.4 \times$, then the same 100\% accuracy \infoclear{is expected} at $1 \times$, and $4.9 \times$ accelerations. Also, for a given signal intensity, note that whenever $100\%$ accuracy was achieved at a particular signal length, all longer signal lengths yielded $100\%$ accuracy as well. For example, for the acceleration factor $1\times$, $100\%$ accuracy was observed for a signal intensity of $1.3$ at a length of $4$ mm. The same held true for all lengths greater than $4$ mm at the same $1.3$ intensity for $1\times$.}
\label{fig:signal_type}
\end{figure}

We adopted a human-label alignment training strategy to ensure that the DLMO achieved performance comparable to human readers on the Rayleigh discrimination task. Training images were selected based on human performance in a series of  two-alternative forced choice (2AFC) studies, conducted by a trained non-physician reader. Each 2AFC study consisted of 50 pairs of singlet and doublet images reconstructed with the rSOS method and characterized by a specific combination of signal intensity, signal length, and acceleration factor. In each trial, the reader was asked to identify the image containing the doublet signal. The 2AFC studies with 100\% \infoclear{(singlet versus doublet) signal classification} accuracy by the reader are shown in Fig.~\ref{fig:signal_type}.

We only included those image sets with signal lengths and intensities where the reader achieved 100\% accuracy when training a DLMO for each acceleration factor.

\section{Sample-size estimation for the human observer two-alternative forced choice study}
We conducted a statistical analysis to determine an appropriate sample size for a 2AFC study aimed at comparing the performance of a DLMO with that of human observers on the Rayleigh task. By specifying the desired significance level and the tolerable difference in performance (e.g., proportion correct) between DLMO and human observers, we explored different combinations of cases ($N_C$), readers ($N_R$), and study designs \infoclear{necessary} to detect the performance difference with sufficient confidence.

Denote human observer proportion correct (PC) by $\theta_0$ and DLMO PC by $\theta_1$. Also, denote similarity margin by $\Delta M$. The two hypotheses for the similarity test are

\begin{equation}
\begin{split}
  H_0: |\theta_0-\theta_1| \ge \Delta M, \\
  H_1: |\theta_0-\theta_1| < \Delta M.
\end{split}
\end{equation}


\subsection{Pilot study}
The process to determine sample size began with a pilot study, which provided initial parameter estimates of the difference in PC and its variance. The pilot study involved three trained human readers. These observers interpreted 50 cases (50 trials or $N_C=50$) in a 2AFC study. The pilot study was fully crossed, which means all cases were read by all readers. From this pilot study, the variance of PC associated with human readers, denoted by $V_0$, was estimated using the iMRMC application \cite{imrmc} to be,
\begin{equation*}
    V_0 = 0.0009179259.
\end{equation*}
This variance can be decomposed into several components. There are multiple ways for the decomposition and we chose the BDG decomposition approach proposed by Gallas et al. \cite{gallas2009framework,gallas2006one,gallas2008reader,gallas2007multireader}. The variance of PC achieved by human readers was decomposed as a scalar product between the coefficients, denoted by vector $\mathbf{c}$, and the moments, denoted by vector $\mathbf{M}$, as follows.
\begin{equation}\label{eq:var}
    V_0 = c_1M_1+c_4M_4+c_5M_5+c_8M_8,
\end{equation} where the moments $\mathbf{M}$ were estimated by iMRMC as shown in Table~\ref{tab:moments} and the coefficients $\mathbf{c}$ are shown in Table~\ref{tab:coeff_pliot_fully_crossed}
\begin{table}[h]
    \caption{Moments of variance components in the pilot study.}
    \begin{center}
    \label{tab:moments}
    \begin{NiceTabular}{c|c|c|c}
	\hline
	\textbf{$M_1$} & \textbf{$M_4$} & \textbf{$M_5$} & \textbf{$M_8$} \\
	\hline
    0.9333 & 0.8712 & 0.9067 & 0.8711\\
    \hline
	\end{NiceTabular}
    \end{center}
\end{table}

\begin{table}[h]
    \caption{Coefficients of variance components in the pilot study.}
    \begin{center}
    \label{tab:coeff_pliot_fully_crossed}
    \begin{NiceTabular}{c|c|c|c}
	\hline
	\textbf{$c_1$} & \textbf{$c_4$} & \textbf{$c_5$} & \textbf{$c_8$} \\
	\hline
    $\frac{1}{N_RN_C}$ & $\frac{N_C-1}{N_RN_C}$ & $\frac{N_R-1}{N_RN_C}$ & $\frac{(N_C-1)(N_R-1)-N_RN_C}{N_RN_C}$\\
    \hline
	\end{NiceTabular}
    \end{center}
\end{table}

The PC achieved by human readers in the pilot study, denoted by $\theta_0$, was
\begin{equation*}
    \theta_0 = 0.9333.
\end{equation*}

The DLMO was trained on 80,000 pairs of images with singlet and doublet signals and tested on a test dataset with $N_C=4000$ cases. Using the iMRMC application, the PC achieved by the DLMO and its variance, denoted by $\theta_1$ and $V_1$, respectively, were as follows,
\begin{equation*}
    \begin{split}
  \theta_1 &= 0.88, \\
  V_1 &= 0.0001188201.
\end{split}
\end{equation*}

Since the samples used to test the DLMO and human observers had no overlap and DLMO was independent from human readers, the standard error of difference in PCs between human readers and DLMO is given by
\begin{equation}
\label{eq:se_diff}
    SE_{diff} = \sqrt{V_0+V_1}.
\end{equation}

\subsection{Sizing}
\begin{figure}[h]
\centering
\includegraphics[width=\linewidth]{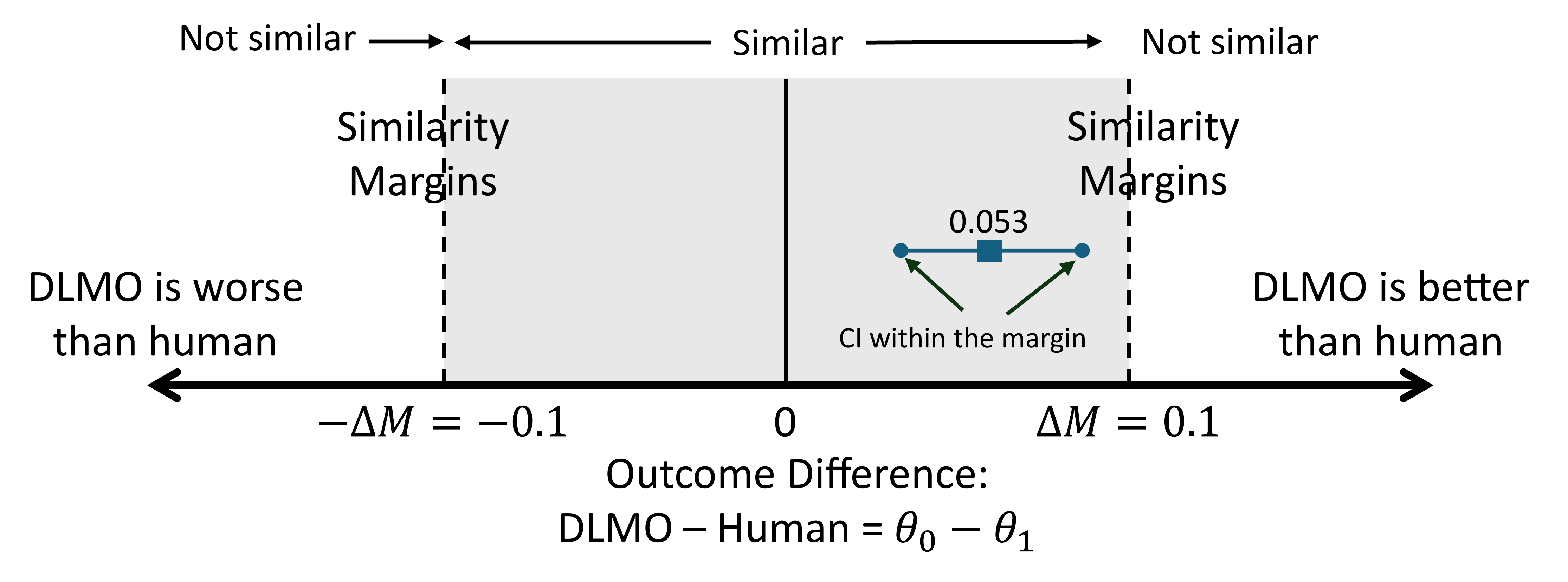}
\caption{An example showing the similarity of a new method (DLMO) to the standard method (human observers).}
\label{fig:similar_diag}
\end{figure}

The difference between PCs from DLMO and human readers was estimated to be $\theta_0-\theta_1 = 0.053$, based on the pilot study. The similarity margin was set to be 0.1, i.e., $\Delta M = 0.1$. In a similarity test, a new method is \infoclear{determined} to be similar to a standard method if the difference in outcomes between the new and standard methods is within similarity margins, as shown in Fig.~\ref{fig:similar_diag}.

The objective of the sample-size determination is to select an appropriate number of cases, readers, and a pivotal study design such that the variance term $V_0$ was small enough for the standard error $SE_{diff}$ in Eq.~\ref{eq:se_diff} to fall below a target standard error. Since there are four statistical tests, we adjusted the two-sided significance level to $0.05/4=0.0125$ \cite{armstrong2014use}, therefore,
\begin{equation*}
    \begin{split}
        Z_{0.0125/2} &= 2.5
    \end{split}
\end{equation*}
Thus, in the pivotal study, the standard error of difference in PCs \infoclear{is expected to be} less than
\begin{equation*}
    \frac{\Delta M - \left(\theta_0-\theta_1\right)}{Z_{\alpha/2}}=0.01868.
\end{equation*}
In this study, we targeted a more conservative standard error of 0.0155, which would allow ample precision for detecting a similarity between DLMO and human readers within a margin of 0.1 at a two-sided significance level of 0.0125. Given the limited number of readers and the time readers could invest, we chose a split-plot design for the 2AFC study \cite{obuchowski1995multireader}, where readers do not read the entire set of test samples but instead read only a subset. Each case is read by only one reader and each reader reads a different, non-overlapped subset of the same number of cases. The coefficients of moments of variance in a 2AFC study with this design are as shown in Table~\ref{tab:coefficients} \cite{gallas2007multireader}.

\begin{table}[h]
    \caption{Coefficients of variance components for split-plot reader design.}
    \begin{center}
    \label{tab:coefficients}
    \begin{NiceTabular}{c|c|c|c}
	\hline
	\textbf{$c_1$} & \textbf{$c_4$} & \textbf{$c_5$} & \textbf{$c_8$} \\
	\hline
     $1 / N_C$ & $(N_C - N_R) / N_R N_C$ &  $0$  & $-1 / N_R$\\
    \hline
	\end{NiceTabular}
    \end{center}
\end{table}

\begin{figure}[h]
\centering
\includegraphics[width=\linewidth]{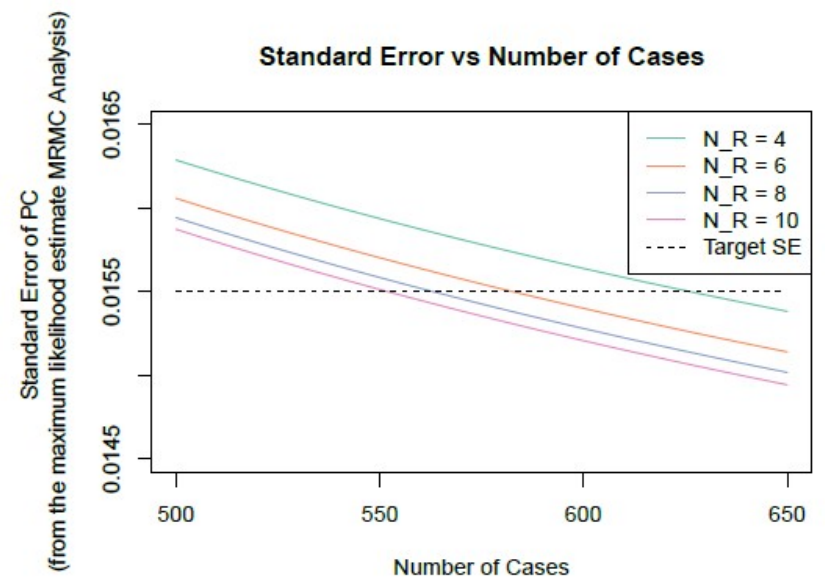}
\caption{The standard error of difference in PCs between DLMO and human readers, with different numbers of readers and cases.}
\label{fig:power_analysis}
\end{figure}

Under this setup, the standard error of difference in PCs between DLMO and human readers can be estimated as a function of $N_C$ (number of cases) for different $N_R$ (number of readers) using Eq~\ref{eq:var} and the moment values in Table~\ref{tab:moments}, as shown in Fig.~\ref{fig:power_analysis}.

\begin{figure}[h]
\centering
\includegraphics[width=0.8\linewidth]{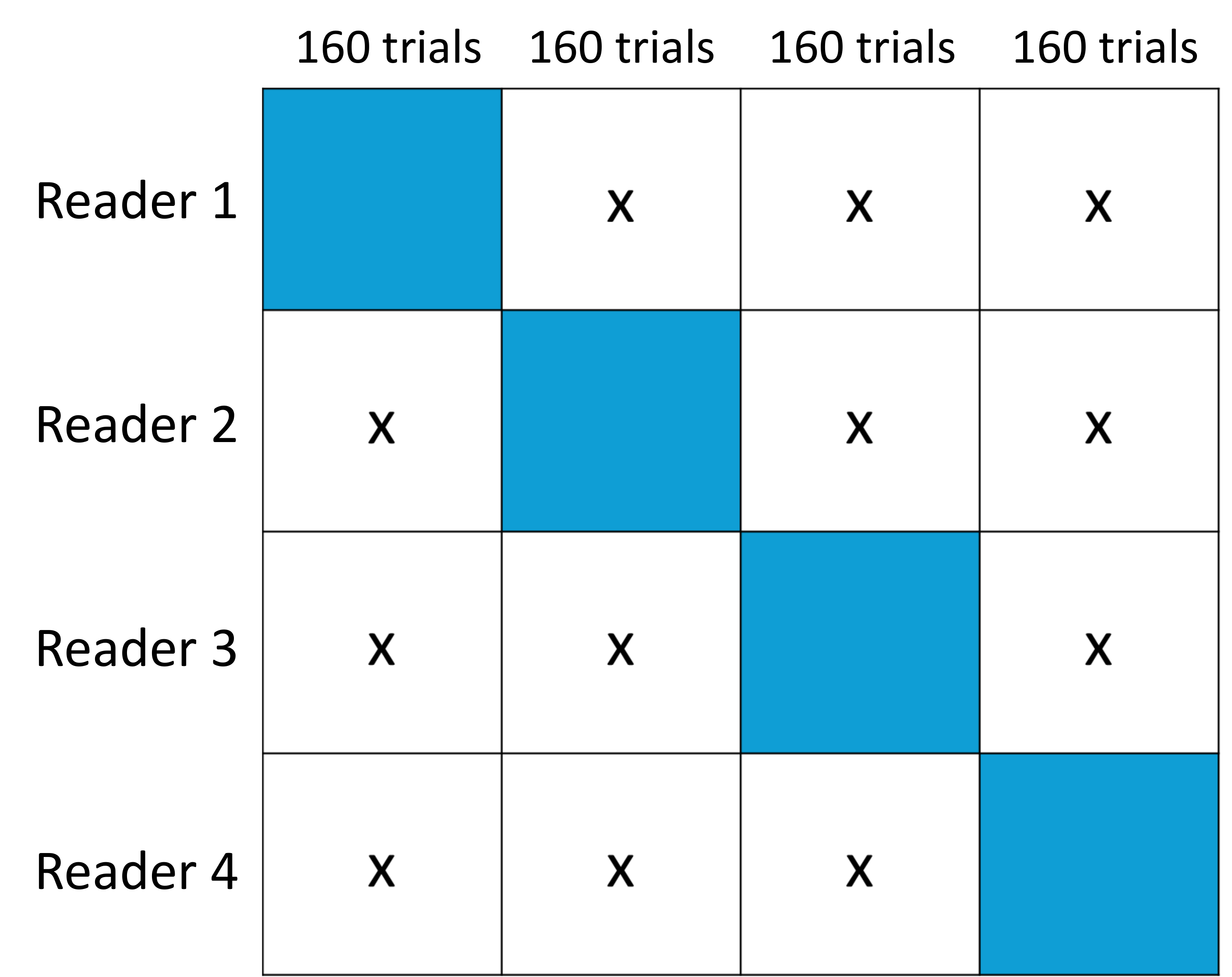}
\caption{Layout of the split-plot reading design used in the pivotal study for each 2AFC-based testing scenario. Blue blocks indicate the image sets assigned to each reader, with each block containing 160 pairs of singlet and doublet samples. Crosses indicate no readings.}
\label{fig:pivotal}
\end{figure}

The curves in Fig.~\ref{fig:power_analysis} indicates that standard error with $N_R = 4$ fell below the target standard error after approximately $N_C = 630$, which was then conservatively increased to 640 in our reader study. This revealed that for the considered margin, to show the similarity of DLMO to human readers with a significance level of 0.05, 4 readers would \infoclear{have} to read at least 640 cases in a flat split-plot 2AFC study without overlap in cases read by readers. This means that each reader would \infoclear{have} to read 160 cases for each 2AFC-based testing scenario. In the main paper, we compared two different reconstruction methods at two different acceleration factors, so there were four testing scenarios. Therefore, 
each reader ended up evaluating $2\times2\times160=640$ singlet and doublet image pairs. This study design is visualized in Fig.~\ref{fig:pivotal}.

\bibliographystyle{IEEEtran}
\bibliography{IEEEabrv_supp}